\documentclass[bibyear]{aa}  

\usepackage{graphicx}
\usepackage[varg]{txfonts}
\usepackage{placeins}           
\usepackage{natbib}
\bibpunct{(}{)}{;}{a}{}{,}
\usepackage[version=4]{mhchem}
\usepackage{booktabs}
\usepackage{amsmath}
\usepackage{hyperref}

\begin{document}

   \title{Formation of the glycine isomer glycolamide (NH$_2$C(O)CH$_2$OH) on the surfaces of interstellar ice grains: Insights from atomistic simulations}

   \author{J. Perrero
          \inst{1,2,3}
          \and
          S. Alessandrini\inst{4}
          \and
          H. Ye\inst{4}
          \and
          C. Puzzarini\inst{4}
          \and
          A. Rimola\inst{1}
          }
   \institute{Department de Química, Universitat Autònoma de Barcelona, Bellaterra, E-08193, Catalonia, Spain.
   \and
   Università degli Studi di Perugia, Dipartimento di Chimica, Biologia e Biotecnologie, Via dell'Elce di Sotto 8, Perugia, I-06123, Italy.
   \and
   Università degli Studi di Torino, Dipartimento di Chimica, Via Pietro Giuria 7, Torino, I-10125, Italy.
    \and
    Department of Chemistry ``Giacomo Ciamician'', University of Bologna, Via F. Selmi 2, Bologna, I-40126, Italy.\\
 \email{jessica.perrero@unito.it, silvia.alessandrini7@unibo.it}
    }         
   \date{Received Month day, 2025; accepted Month day, 2025}

\abstract
  
   {Syn-glycolamide, a glycine isomer, has recently been detected in the G+0.693-0.027 molecular cloud. Investigations on its formation in the interstellar medium could offer insights into synthetic routes leading to glycine in prebiotic environments.}
   {Quantum chemical simulations on gly\-col\-amide (\ce{NH2\-C(O)\-CH2\-OH}) formation on interstellar ice mantles, mimicked by a water ice cluster model, are presented.}
   {Glycolamide synthesis has been here modeled considering a stepwise process: the coupling between formaldehyde (\ce{H2CO}) and the radical of formamide (\ce{NH2CO^.}) occurs first, forming the glycolamide precursor \ce{NH2C(=O)CH2O^.}, which is then hydrogenated to give anti-glycolamide. We hypothesize that anti-to-syn interconversion will occur in conjunction with glycolamide desorption from the ice surface.}
   {The reaction barrier for \ce{NH2\-C(O)\-CH2O^.} formation varies from 9 to 26 kJ mol$^{-1}$, depending on surface binding sites. Kinetic studies indicate that this reaction step is feasible in environments with a $T > 35~\text{K}$, until desorption of the reactants. The hydrogenation step leading to anti-glycolamide presents almost no energy barrier due to the easy H atom diffusion towards the \ce{NH2C(O)CH2O^.} intermediate. However, it competes with the extraction of an H atom from the formyl group of \ce{NH2C(O)CH2O^.}, which leads to formyl formamide, \ce{NH2C(O)CHO}, and \ce{H2}. Nonetheless, according to our results, anti-glycolamide formation is predicted to be the most favored reactive channel.}
  {}

\titlerunning{Formation of glycolamide on interstellar grains}
\authorrunning{J. Perrero et al.}

\maketitle
%

\section{Introduction}

The core idea behind the different scenarios suggested for the origin of life on Earth is that simple biomolecular building blocks as amino acids, sugars and nucleobases evolved into polymers and macromolecules in the early Earth environment \citep{saladino2012}.  
What kind of polymers started the self-replication process is still an open question and, to understand how these species could have formed, several points have to be addressed, among them the origin of the early organic material. Such molecules might have been either delivered to Earth by comets and meteorites or incorporated during the planet formation, but their synthesis occurred in the interstellar medium (ISM). The ISM is a surprisingly chemically rich environment with a census of about 330 species detected to date \citep{Mul01,Mul05,McGuire_2022}. Several of them show a strong prebiotic character, such as urea (\ce{NH2C(O)NH2}) \citep{belloche2019} and ethanolamine (\ce{OHCH2CH2NH2}) \citep{rivilla2021discovery}, and their number is expected to grow due to the continuous advances in observational facilities and their sensitivity.

Recently, the first observation of glycolamide (\ce{NH2C(O)CH2OH}, also known as 2-hydroxyacetamide) was reported in the G+0.693-0.027 molecular cloud \citep{rivilla2023first}. This species has two conformers, depending whether the hydroxyl (\ce{-OH}) moiety is \textit{syn} or \textit{anti} with respect to the
carbonyl group (\ce{-C=O}). Both conformers were considered in the observation toward the G+0.693-0.027 molecular cloud, but only the \textit{syn} conformer was detected, which is the most stable species in the gas phase by $\sim$ 4 kJ mol$^{-1}$ according to experimental data \citep{maris2004conformational}. The increased stability of the \textit{syn} conformer is due to an intramolecular hydrogen bond between the hydroxyl and carbonyl moieties, which is substituted by a weaker H-bond interaction between the \ce{NH2} group and the oxygen of the OH moiety in the \textit{anti} conformation. 

Glycolamide contains both an amide functional group (\ce{O=C\-NH2}) and a hydroxyl group (\ce{OH}) in $\alpha$ position, closely resembling an amino acid. The detection of glycolamide is important because it is a structural isomer of the non-chiral amino acid glycine, which has been identified in the 81P/Wild 2 \citep{Elsila2009} and 67P/C-G \citep{Altwegg2016} comets as well as in various carbonaceous meteorites \citep{pizzarello2006chemistry,burton2012understanding,hiroi2001tagish}. Several experimental studies have suggested that glycine could form without the need of energetic irradiation (such as UV photons and cosmic rays) in interstellar water-rich ices \citep{ioppolo2021non}. However, despite numerous attempts, the presence of glycine in the ISM has not been unambiguously confirmed yet \citep[e.g.,][]{lattelais2011detectability,hollis2003sensitive}. Therefore, investigating formation pathways of its isomers, such as glycolamide, can be useful in providing new insights into its synthesis. Moreover, the prebiotic role of glycolamide does not end with its chemical relation to glycine; indeed, this species might have played a pivotal role in the early Earth conditions, being able to undergo polycondensation reactions that lead to dipeptides, such as N-(2-amino-2-oxoethyl)-2-hydroxyacetamide. These processes could represent a first step into the formation of peptide chains \citep{szHori2011chemical}.

In addition to \textit{syn}-glycolamide, few other amides have been detected and identified in the ISM, i.e., urea (\ce{NH2C(O)NH2}), acetamide (\ce{CH3C(O)NH2}), N-methylformamide (\ce{HC(O)NHCH3}) and formamide (\ce{HC(O)NH2}) \citep{belloche2017,belloche2019,jimenez2020,ligterink2022,zeng2023}. This latter is considered the precursor of the amides family, but other chemical links have been suggested as relevant. Indeed, formation of amides in the ISM can arise from isocyanates (\ce{R-NCO}) \citep{colzi2021} as their hydrogenation can lead to amide precursors. For example, the hydrogenation of \ce{HNCO} can form the \ce{NH2CO^.} formamide radical, which in turn has been suggested by quantum chemical computations to be a key intermediate of the \ce{^.CN} + \ce{H2O} and \ce{^.NH2} + \ce{CO} reactions on interstellar water ice mantles \citep{rimola2018,silva2024pathways}. For this reason, many studies on the formation of amides in the ISM have focused on their ice-surface synthesis. For example, astrochemical models suggest that formamide could be formed on ices starting from \ce{^.CN/HCN} and \ce{H2O}, while acetamide could result from the addition of \ce{^.CH3} to the \ce{^.NH2CO} radical \citep{garrod2008complex}. In both processes, the role of energetic photons was also investigated. It was found that formamide is obtained from irradiation of \ce{CO}:\ce{NH3} ices \citep{jones2011mechanistical}, while acetamide is produced in ices composed by \ce{NH3} and \ce{CH3CHO} \citep{marks2023prebiotic}. Instead, mixtures of \ce{CH3NH2} and \ce{CO} ices were suggested as possible precursors of N-methylformamide \citep{bossa2012solid,frigge2018vacuum}. However, not much is known about the formation of glycolamide on ices. 

Interstellar ices have a versatile role in astrochemistry. They can act as reactants concentrators, allowing species to reside in close proximity and thus favoring their reaction. The ice surfaces can also play a chemical catalytic effect, providing alternative reaction mechanisms with lower energy barriers compared to the analogous gas-phase processes. Finally, the icy mantles can exert a third-body effect that dissipates the energy liberated by exothermic processes, without harming the products so formed. Accordingly, ice surface reactions are particularly relevant for the G+0.693-0.027 molecular cloud, where glycolamide and the other amides have been detected. This region is thought to undergo a cloud-cloud collision and, as a consequence, large-scale shocks occur. These events sputter the dust grains and therefore many species are expected to be released into the gas phase \citep{zeng2020cloud,jimenez2008parametrization}. 

The aim of the present study is to investigate a plausible formation pathway for glycolamide, in which the radical \ce{NH2CO^.} (formed by either hydrogenation of HNCO, or reaction of \ce{^.CN} with icy \ce{H2O}, or both) acts as its precursor by reacting with \ce{H2CO} (an available species within interstellar ices). The overall formation pathway considered here is composed of two steps that can be sketched as follows:
\vspace{-1ex}
\begin{equation}\label{eq:first}
   \mathrm{H_2CO} + \mathrm{NH_2CO^{\bullet}} \rightarrow \mathrm{NH_2C(O)CH_2O^{\bullet}}
\end{equation}
\vspace{-2ex}
\begin{equation}\label{eq:second}
   \mathrm{NH_2C(O)CH_2O^{\bullet}} + \mathrm{H^{\bullet}} \rightarrow \mathrm{NH_2C(O)CH_2OH}
\end{equation}

In this work, we investigate the glycolamide synthetic route taking place on the surfaces of interstellar water ice grains by means of state-of-the-art quantum chemical simulations and mimicking the water ice surfaces with an atomistic 18-\ce{H2O} cluster model. To this end, the potential energy surfaces (PESs) of the two reaction steps shown above have been explored in order to obtain their intrinsic energetics. The study has then been complemented with a kinetic analysis to predict their feasibility under interstellar conditions.   

The paper is organized as follows. The next section illustrates the computational methodology employed to investigate steps (\ref{eq:first}) and (\ref{eq:second}). Section \ref{results} provides an overview of the results, followed by their discussion in Section \ref{discussion}. Finally, the main conclusions are summarized in Section \ref{conclu}. 

\section{Computational Methodology}\label{methods}
All the computations carried out in this work (namely, geometry optimizations, frequency calculations, and single-point calculations for energy refinement) were performed using the ORCA 5.0.4 software \citep{neese2022}.
The strategy employed to understand the reaction mechanism and to establish the most suitable computational methodology can be summarized as follows.

The gas-phase reaction steps (\ref{eq:first}) and (\ref{eq:second}) (i.e., in the absence of the water ice model) were characterized to obtain their intrinsic energetics (that is, energy barriers and reaction energies) and to benchmark the performance of different cost-effective quantum chemical methods grounded on the density functional theory (DFT). 
The final goal was to identify the most appropriate functional to describe the formation of glycolamide on the water ice cluster model, as these calculations are more expensive in terms of computational cost than those describing gas-phase reactions.
The density functionals tested in this preliminary benchmarking study were B3LYP \citep{lee1988development,becke:1988,becke1992density}, BHLYP \citep{bhandhlyp-becke1993,lee1988development}, M062X \citep{m062x-zhao}, and $\omega$B97X \citep{wb97x-chai}. Since these do not properly cope with the dispersion interactions, they were all corrected with the Grimme's D3 zero damping dispersion or, when available, the D3(BJ) Becke-Johnson dispersion terms \citep{grimme:2010,grimme:2011}. Additionally, the $\omega$B97M-V functional \citep{mardirossian2016omegab97m}, a van der Waals density functional that employs the same electron density to incorporate dispersion effects, was tested.
Each functional was combined with the ma-def2-TZVP basis set \citep{zheng2011}. All these methods were used to optimize the geometry of the pre-reactant complex (given by the gas-phase adduct between \ce{H2CO} and \ce{NH2CO^.}), the transition state (TS) and the intermediate (\ce{NH2COCH2O^.}), as well as the TS for the hydrogenation of the latter yielding glycolamide (\ce{NH2COCH2OH}). 

\begin{figure}[tb]
    \centering
    \includegraphics[width=0.90\columnwidth]{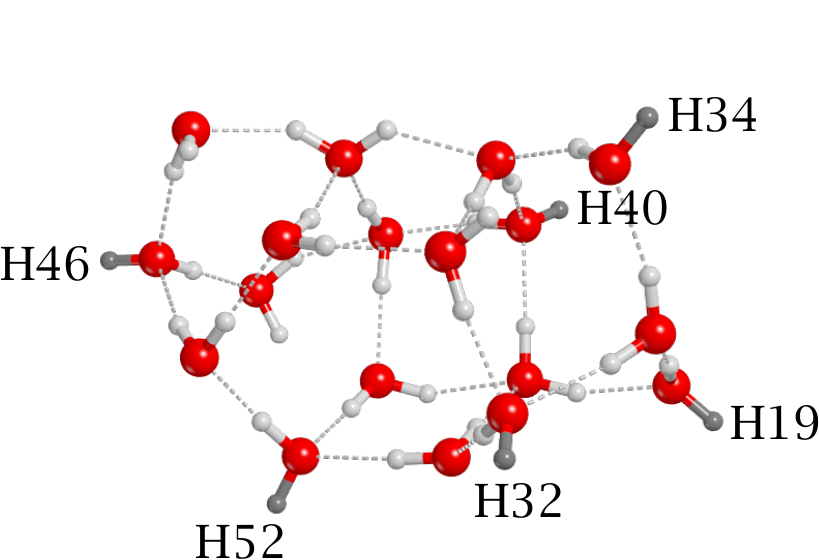}
    \caption{The 18 water molecules cluster (W18) used in this work. Oxygen atoms are shown in red, while hydrogens are in light gray. The dangling hydrogen (dH) atoms selected to adsorb \ce{H2CO} and \ce{NH2CO^.} are highlighted in dark gray. }
    \label{fig:ice}
\end{figure}

To assess the accuracy of the above-mentioned methods, we performed single-point energy calculations onto the DFT optimized structures using the CCSD(T)-F12 method, which is the F12 explicitly correlated variant of the coupled cluster theory with singles, doubles and a perturbative treatment of triples \citep[][and references therein]{adler2007}. Currently, this is considered as the gold standard in terms of accuracy. However, it is dramatically more expensive than DFT approaches, making it almost unpractical for large systems. CCSD(T)-F12 calculations require the use of correlation consistent basis sets. In our case, cc-pVTZ-F12 was employed as the main basis set, with cc-pVTZ-F12-CABS, aug-cc-pVTZ, and aug-cc-pVQZ/C as auxiliary basis sets \citep{peterson2008systematically,dunning1989gaussian,kendall1992electron}. The comparison of the DFT results with those obtained at the CCSD(T)-F12 level allowed us to establish the most suitable functional to characterize steps (\ref{eq:first}) and (\ref{eq:second}) in the presence of the water ice cluster. At this stage, we also computed the PES of the gas-phase isomerization from \textit{anti-} to \textit{syn-}glycolamide at the revDSD-PBEP86/jun-cc-pVTZ level of theory \citep[briefly, revDSD/junTZ,][]{papajak2011convergent,santra2019minimally} and compared it with the best functional resulting from the benchmark study.

Afterwards, we proceeded with the study of the same reactions on the ice surface. For this purpose, an 18 \ce{H2O} molecules cluster model (W18) was used. This represents a compact, amorphous, and flat water ice surface (see Figure \ref{fig:ice}), which was adopted in previous studies \citep{rimola2014,rimola2018,enrique-romero2019,perrero2024synthesis}. This model is a compromise between an extended structure, closer to reality, and a smaller cluster comprising fewer units, allowing for highly accurate calculations.

Steps (\ref{eq:first}) and (\ref{eq:second}) were simulated adopting a Langmuir-Hinshelwood mechanism; thus, the first step consisted in the adsorption of the reactants, followed by their reaction. \ce{H2CO} and \ce{NH2CO^.} were adsorbed singularly on six adsorption sites represented by dangling hydrogen (dH) atoms on the W18 cluster (illustrated in Figure \ref{fig:ice}). The species were manually adsorbed based on the H-bonding complementarity between the dH atoms and the donor groups of the reactants.

The strength of the interaction between reactants and each binding site is defined using the corresponding binding energy (BE) which is expressed as the opposite of the interaction energy ($\Delta E$) corrected for the vibrational zero-point energy (ZPE):

\begin{equation}
    \text{BE} = -(\Delta \text{E} + \Delta \text{ZPE})  
\end{equation}

where $\Delta E = E_{\text{complex}} - E_{\text{W18}} - E_{\text{molecule}}$ and $\Delta \text{ZPE} = \text{ZPE}_{\text{complex}} - \text{ZPE}_{\text{W18}}- \text{ZPE}_{\text{molecule}}$. Due to its definition, BE results in a positive quantity for favorable interactions between the adsorbates and the surface.
The ZPE, associated with the residual vibrational motion of a species at a temperature of 0~K, is obtained from frequency calculations, within the harmonic approximation. Furthermore, harmonic frequency calculations also allowed us to confirm the nature of the stationary points (minima or TSs) identified on the PES.  

The adsorption complexes obtained for each of the reactants were then classified from the strongest to the weakest with the aim of identifying the couples of binding sites (i) that grant the largest BE, (ii) whose reactants are in close proximity to promote a reaction. The identified couples were used as the pre-reactant structures to explore the PESs.

Reaction energy barriers were calculated as the energy difference between the TS structure and the pre-reactant complex, while reaction energies as the energy difference between the product and the pre-reactant complex. The addition of the ZPE corrections to these energy values led to enthalpy barriers ($\Delta H(0)_{\text{TS}}$) and reaction enthalpies ($\Delta H(0)_{\text{R}}$) at $T = 0~\text{K}$. 
Once the PESs were characterized, the corresponding energetics relative to the formation of the intermediate \ce{NH2COCH2O^.} was used to analyze its kinetics by means of the semi-classical Eyring's equation:

\begin{equation}\label{eq:eyring}
        k(T) = \kappa \times \frac{k_B T} {h} \times exp (- \Delta G ^\ddagger /RT)
\end{equation}
    
where $\kappa$ is the tunneling transmission coefficient (which, for this reaction, is assumed to be 1 because tunneling effects are considered to be negligible as the reactive atoms are heavy species), $k_B$ is the Boltzmann constant, $T$ is the temperature, $h$ is the Planck constant, $\Delta G ^\ddagger$ is the Gibbs energy barrier and $R$ is the gas constant.   
The $\Delta G ^\ddagger$ values were obtained by applying thermochemical corrections (from statistical thermodynamics formulae) to the energy values in the $T= 5-100$~K temperature range, with a step of 5~K, and in the low pressure limit. The computation of the unimolecular rate constant $k(T)$ from equation (\ref{eq:eyring}) allows determining the temperature-dependent half-life time of the reactants as $t_{1/2}(T) = ln(2)/k(T)$, which is defined as the time for the reactants to be half consumed.

After the formation of \ce{NH2COCH2O^.}, the final step consisted of its hydrogenation via \ce{H^.} addition, thus being a radical-radical recombination. Initially, the H atom was adsorbed on the W18 cluster in the presence of \ce{NH2COCH2O^.}, with the two unpaired electrons having the same electronic spin and thus leading to a triplet system. This electronic state is not reactive (due to the Pauli repulsion principle); its geometrical optimization was useful to obtain a pre-reactant structure before the hydrogenation process. To allow the formation of the \ce{O-H} bond, the system was then re-optimized in the singlet electronic state but as an open-shell biradical system. For this purpose, to correctly  describe its electronic structure, calculations were performed within an unrestricted formalism and adopting the broken-(spin)-symmetry \textit{ansatz}  \citep{neese_2004,abe2013}. This approach allows two unpaired electrons with opposite spins to reside in different orbitals, thus avoiding the need to force the recombination between the two radicals. 
The validity of this strategy was proven in previous literature studies on biradical systems \citep{Enrique-Romero2022}, where it was also compared with CASPT2 calculations \citep{Enrique-romero2020}.
In those cases where the spontaneous formation of \ce{NH2COCH2OH} was not observed, the presence of a diffusion barrier of the H atom on the ice surface was confirmed by performing scan calculations along the \ce{H-O} coordinate with steps of 0.1~\r{A} and tight convergence criteria for the optimization.

\begin{figure*}[htb]
    \centering
    \includegraphics[trim=2cm 1.5cm 2cm 2cm,width=0.90\textwidth]{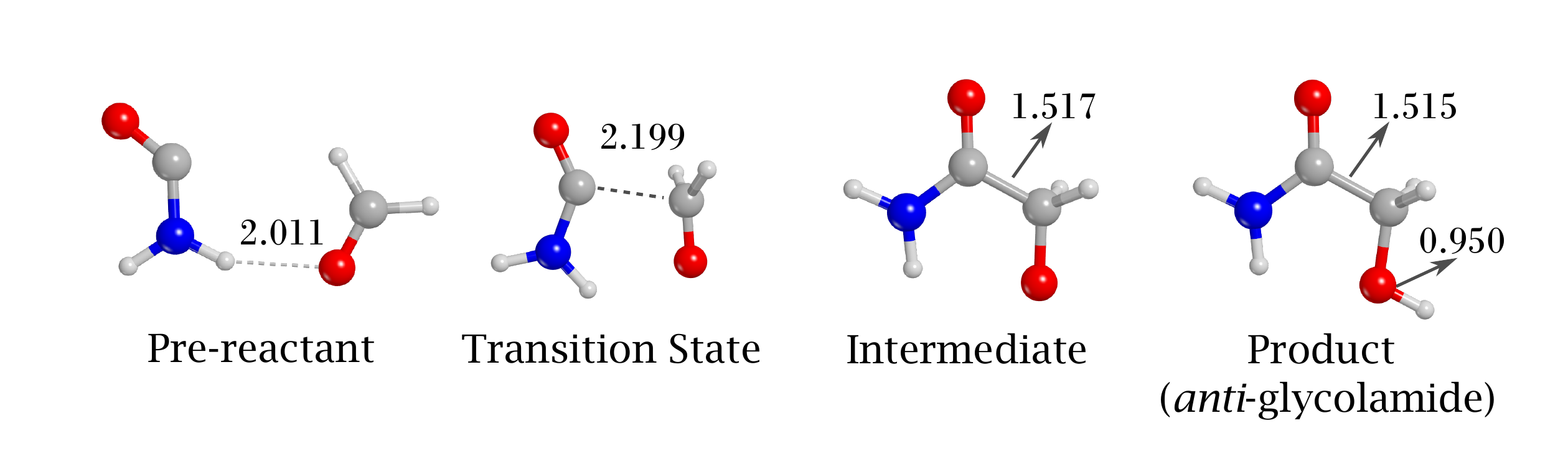}
    \caption{Structures of the stationary points for the gas-phase reaction evaluated at the BHLYP-D3(BJ)/ma-def2-TZVP level. Distances in \r{A}.}
    \label{fig:gas}
\end{figure*}

\section{Results}\label{results}

\subsection{Gas-phase reactions}

\begin{table*}[!tb]
\centering
\caption{Energetics of the gas-phase formation of glycolaldehyde. Energies are in kJ mol$^{-1}$. 
A visual representation of these data is available in Figure \ref{benchmark_plot} of the Appendix.}\label{tab:benchmark}
\setlength{\tabcolsep}{10pt}
\begin{tabular}{cccccccccc}
\hline\hline
\\[-2.0ex]
                           & \multicolumn{6}{c}{Step (\ref{eq:first})} & \multicolumn{3}{c}{Step (\ref{eq:second})}  \\ \cline{2-7} \cline{8-10} \\[-2.0ex]
                           & \multicolumn{3}{c}{Energy barrier}        & \multicolumn{3}{c}{Reaction energy}  & \multicolumn{3}{c}{Reaction energy}  \\ \\[-2.0ex]
DFT geom.\tablefootmark{a} &  DFT\tablefootmark{b} &  SP\tablefootmark{c}  & err.\tablefootmark{d} & DFT\tablefootmark{b} & SP\tablefootmark{c} & err.\tablefootmark{d} & DFT\tablefootmark{b} & SP\tablefootmark{c} & err.\tablefootmark{d} \\ 
\hline
\\
B3LYP-D3(BJ)	    &     2.0 &  12.9 & 84\% & -29.8 & -20.9 & 43\% & -441.9 & -472.8 & 7\% \\
BHLYP-D3(BJ)	    &	 13.1 &  14.3 &  9\% & -48.4 & -25.3 & 91\% & -439.8 & -469.3 & 6\% \\
M062X-D3	        &     5.8 &  13.0 & 55\% & -33.9 & -25.3 & 34\% & -463.9 & -469.7 & 1\% \\
$\omega$B97M-V	    &     7.1 &  14.1 & 50\% & -35.1 & -24.3 & 44\% & -459.2 & -469.6 & 2\% \\
$\omega$B97X-D3(BJ)	&     6.6 &  13.8 & 52\% & -38.5 & -24.2 & 59\% & -452.0 & -469.6 & 4\% \\
\hline\hline
\end{tabular}
\tablefoot{
\tablefoottext{a}{``DFT geom.'' indicates the DFT level at which the geometries were optimized. In all cases, the ma-def2-TZVP basis set was employed.}
\tablefoottext{b}{``DFT'' indicates that the energy was evaluated at the same level as the ``DFT geom.''.} 
\tablefoottext{c}{``SP'' refers to single-point CCSD(T)-F12/cc-pVTZ-F12 energy evaluations on top of the ``DFT geom.''.} 
\tablefoottext{d}{``err.'' denotes the absolute relative deviation of ``DFT'' with respect to ``SP''.}}
\end{table*}

As mentioned above, the first step of this investigation consists of a benchmarking study using the gas-phase reactivity. The structures of stationary points on the gas-phase reactive PES (namely, pre-reactant, TS, intermediate, and product) are shown in Figure \ref{fig:gas}. While the reaction step (\ref{eq:first}) shows an energy barrier, this is not the case for the reaction step (\ref{eq:second}) because hydrogenation of the \ce{NH2COCH2O^.} intermediate occurs spontaneously (the \ce{O-H} bond directly forms upon geometry optimization). The energy barriers and reaction energies (without ZPE corrections) calculated for each step are summarized in Table \ref{tab:benchmark}. In this table, the results for the DFT functionals considered are reported and compared with the CCSD(T)-F12 counterparts. 

The energy barriers calculated at the CCSD(T)-F12 level on top of the different DFT geometries lie around 13-14 kJ mol$^{-1}$, indicating a robust value irrespective of small geometry variations. In contrast, different energy barriers --ranging between 2 and 13 kJ mol$^{-1}$-- are obtained when using different DFT functionals, overall showing that DFT tends to underestimate the energy barrier. 
The relative error given by each functional with respect to CCSD(T)-F12 has been worked out as an indicator of the accuracy. Table \ref{tab:benchmark} points out that BHLYP-D3(BJ) is the most suitable functional for the description of the energy barrier of step (\ref{eq:first}), with a relative error of 9\%. At the same time, though, it provides the worst performance in describing its exoergicity, the corresponding value being almost twice as that estimated by CCSD(T)-F12. However, the parameter that determines whether the reaction does or does not proceed towards the formation of the product is the energy barrier. Moreover, an overestimation of the energy released by the formation of the product does not affect the outcome of the reaction. In fact, the reaction is exoergic at all the levels of theory considered, thus indicating that product formation is thermodynamically favored in any case.
Additionally, the benchmark of step (\ref{eq:second}) shows that the error of all functionals on its exoergicity is comprised between 1\% and 7\%. A similar argument as that discussed before applies to this energy difference. 
For these reasons, the BHLYP-D3(BJ)/ma-def2-TZVP level of theory was selected for the characterization of glycolamide formation on the W18 cluster.

\begin{figure}[htb]
    \centering
    \includegraphics[width=0.99\columnwidth]{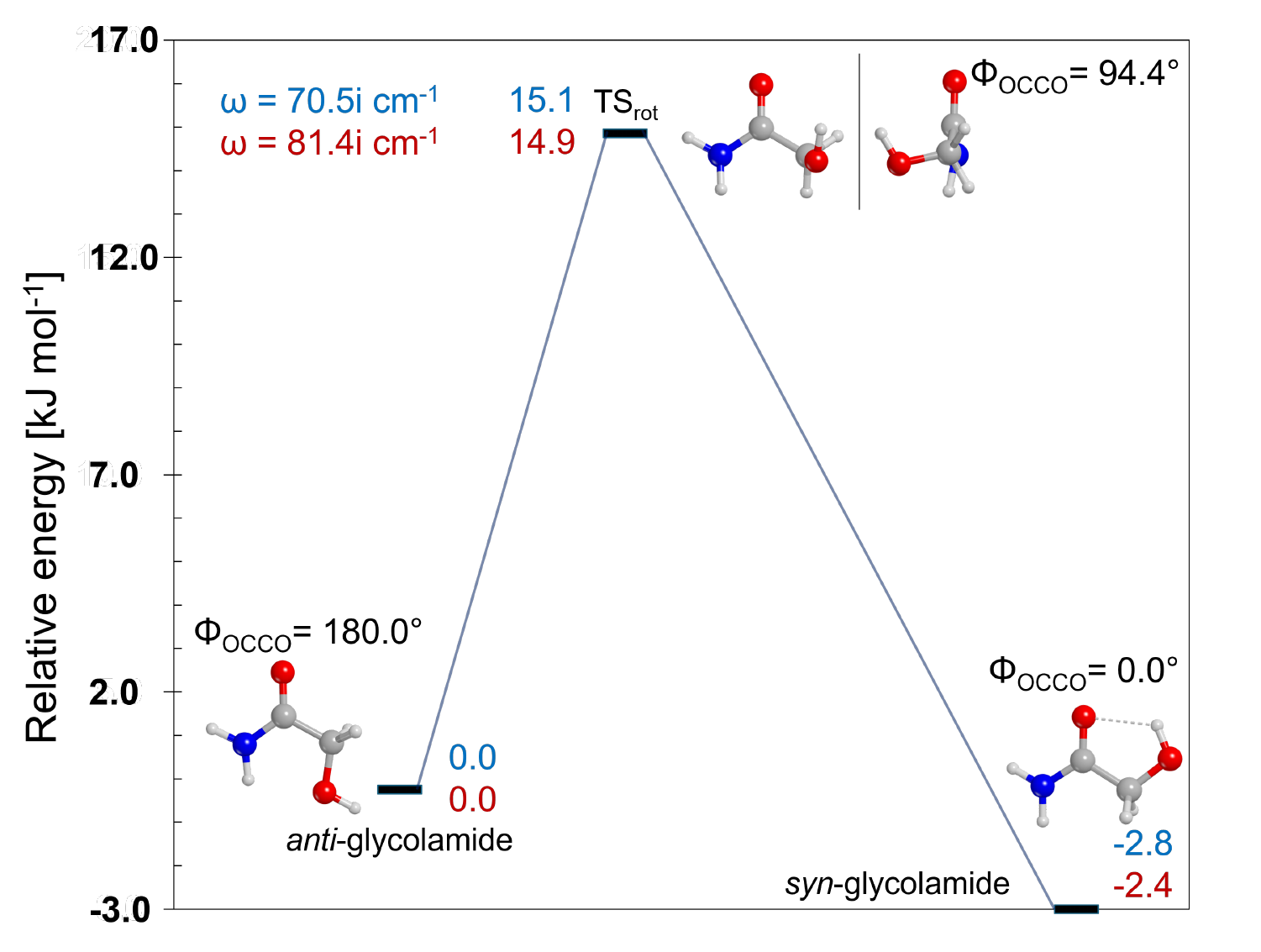}
    \caption{ZPE-corrected gas-phase PES (in kJ mol$^{-1}$) of the interconversion between the \textit{anti} and \textit{syn} conformers of glycolamide, calculated at the BHLYP-D3(BJ)/ma-def2-TZVP (in red) and revDSD/junTZ (in blue) levels of theory. The dihedral angle $\Phi_{OCCO}$, defining the \textit{syn} and \textit{anti} conformations and adopted as the reaction coordinate that drives the interconversion process, are also shown for the localized stationary points (in degrees). For each level of theroy, the TS imaginary frequency ($\omega$, in cm$^{-1}$) is also reported. For the TS, two different orientations are shown for the sake of clarity.}
    \label{fig:synanti}
\end{figure}

As it can be noted in Figure \ref{fig:gas}, the final product obtained from the reaction of \ce{NH2CO} with \ce{H2CO} is \textit{anti}-glycolamide. This is because the two reactants maximize their interaction by forming two H-bonds that orient the transition state only towards the \textit{anti} conformation. Since only \textit{syn}-glycolamide was observed in the ISM, we characterized the interconversion barrier between the two conformers in the gas phase, as shown in Figure \ref{fig:synanti}. The geometries and their relative energies were computed using the BHLYP-D3(BJ)/ma-def2-TZVP and revDSD/junTZ levels of theory. Isomerization and its barrier ($\sim$15 kJ mol$^{-1}$) will be discussed in detail in Section \ref{discussion}. 

\subsection{Adsorption of \texorpdfstring{\ce{NH2CO^.}}{NH2CO} and \texorpdfstring{\ce{H2CO}}{H2CO} on W18}

The first step to simulate the formation of glycolamide on the W18 cluster is the adsorption of \ce{H2CO} and \ce{NH2CO^.} on the six binding sites identified in the ice model (see Figure \ref{fig:ice}). This resulted in 12 adsorption complexes for \ce{H2CO} and 16 for \ce{NH2CO^.}. This variety arises from the different orientation that each species can adopt upon adsorption on the same binding site, which in turn depends on the interacting neighboring atoms. The BEs obtained for the \ce{H2CO} and \ce{NH2CO^.} adsorption complexes are listed in Table \ref{tab:be}. 

Both \ce{H2CO} and \ce{NH2CO^.} are polar species that interact with the water ice mainly through H-bonds. \ce{H2CO} spans a range of BE = 15.7$-$27.1 kJ mol$^{-1}$, which is governed by both the strength of the H-bond accepted by the carbonyl moiety and the secondary interaction between the \ce{CH2} group and the surrounding water molecules. A shorter H-bond length is indicative of a stronger interaction.
In contrast, the BEs of \ce{NH2CO^.} cover the range of  21.7$-$51.8 kJ mol$^{-1}$. The interaction of \ce{NH2CO^.} with the cluster occurs through two H-bonds in the majority of the cases, with the exception of a few cases in which a third longer H-bond is established (such as 40c and 52c). The binding site 34a shows the smallest BE because \ce{NH2CO^.} interacts with only one water molecules, while the largest BE corresponds to the complex 32c, in which two short and well-oriented H-bonds are formed. For clarity, the geometries of the adsorbed reactants are shown in sections \ref{H2COad} and \ref{NH2COad} of the Appendix.

\begin{table}[t!]
\centering
\caption{Interaction energies ($\Delta$E) and ZPE-corrected binding energies (BE), in kJ mol$^{-1}$, for \ce{H2CO} and \ce{NH2CO^.} adsorbed on the W18 cluster.$^a$}\label{tab:be}
\begin{tabular}{@{}ccc ccc}
\toprule
\multicolumn{3}{c}{\ce{H2CO}} &  \multicolumn{3}{c}{\ce{NH2CO^.}} \\
\midrule
 Binding Site\tablefootmark{a} & $\Delta$E & BE & Binding Site\tablefootmark{a} & $\Delta$E & BE \\
\midrule
19a	& -27.1	& 15.7 & 19a & -39.0	& 33.6 \\
19b & -27.2 & 16.5 & 19b & -44.5	& 36.9 \\
32a	& -30.1	& 23.2 & 32a & -48.5	& 41.3 \\
32b & -30.9 & 22.0 & 32b & -50.0	& 43.7 \\
32c & -32.0 & 25.7 & 32c & -60.9	& 51.8 \\
34a	& -26.5	& 19.5 & 34a & -27.5	& 21.7 \\
34b & -31.8 & 25.8 & 34b & -38.8	& 33.0 \\
    &       &      & 34c & -48.6	& 40.9 \\
40a	& -39.8	& 27.1 & 40a & -46.2	& 39.8 \\
    &       &      & 40b & -47.6	& 41.2 \\
    &       &      & 40c & -44.2    & 38.0 \\
46a	& -31.6 & 24.2 & 46a & -53.5	& 47.0 \\
46b & -30.4 & 23.8 & 46b & -47.5	& 39.4 \\
46c & -34.0 & 23.8 &     &          &      \\
52a	& -31.8	& 21.2 & 52a & -50.6	& 42.6 \\
    &       &      & 52b & -48.7	& 40.8 \\
    &       &      & 52c & -49.8	& 40.6 \\
\bottomrule
\end{tabular}
\tablefoot{The geometries of the different adsorption complexes are available in the Appendix.
\tablefoottext{a}{The labels a, b, and c refer to different adsorption complexes obtained at the same binding site.}
}
\end{table}

In order to determine a suitable combination of binding sites to derive a pre-reactive complex for the reaction between \ce{H2CO} and \ce{NH2CO^.}, we established two conditions to be fulfilled: (i) the strongest binding sites for each adsorbate are preferred, since they determine the positions in which the species are more favorably adsorbed; (ii) reactants need to be close one to each other, at a distance that allows their reaction without the need of prior diffusion on the ice surface. 
According to these criteria, we identified four couples, resulting from the combination of the most strongly bounded adducts for both \ce{H2CO} and \ce{NH2CO^.}. The derived structures were optimized and ordered according to the final BE of the reactants in Table \ref{tab:reaction}. Due to the simultaneous presence of both species on the W18 cluster and the consequent structural rearrangement of their atomic positions compared to their isolated complexes, the BEs obtained for these structures are not equal to the sum of the isolated \ce{H2CO} and \ce{NH2CO^.} BEs, but they are larger. This behavior is explained by the additional interactions established between the two adsorbates.

\subsection{Reactivity between \texorpdfstring{\ce{NH2CO^.}}{NH2CO} and \texorpdfstring{\ce{H2CO}}{H2CO} on W18}

For each of the complexes listed in Table \ref{tab:reaction}, we identified the TS connected to the formation of \ce{NH2COCH2O^.}. As expected, the reaction mechanism is analogous to the one characterized in the gas-phase process, that is, a \ce{C-C} bond forms after the two reactants face each other. However, the calculated energy barriers from the different ice-surface reactants are different, spanning the $\Delta$H(0)$_{TS}$ = 9.6 -- 26.0~kJ mol$^{-1}$ range. The complete summary of the structures for these reactions is reported in Figure \ref{tabrection_SI} of the Appendix. 

\begin{figure*}[htb]
    \centering
    \includegraphics[width=0.95\linewidth]{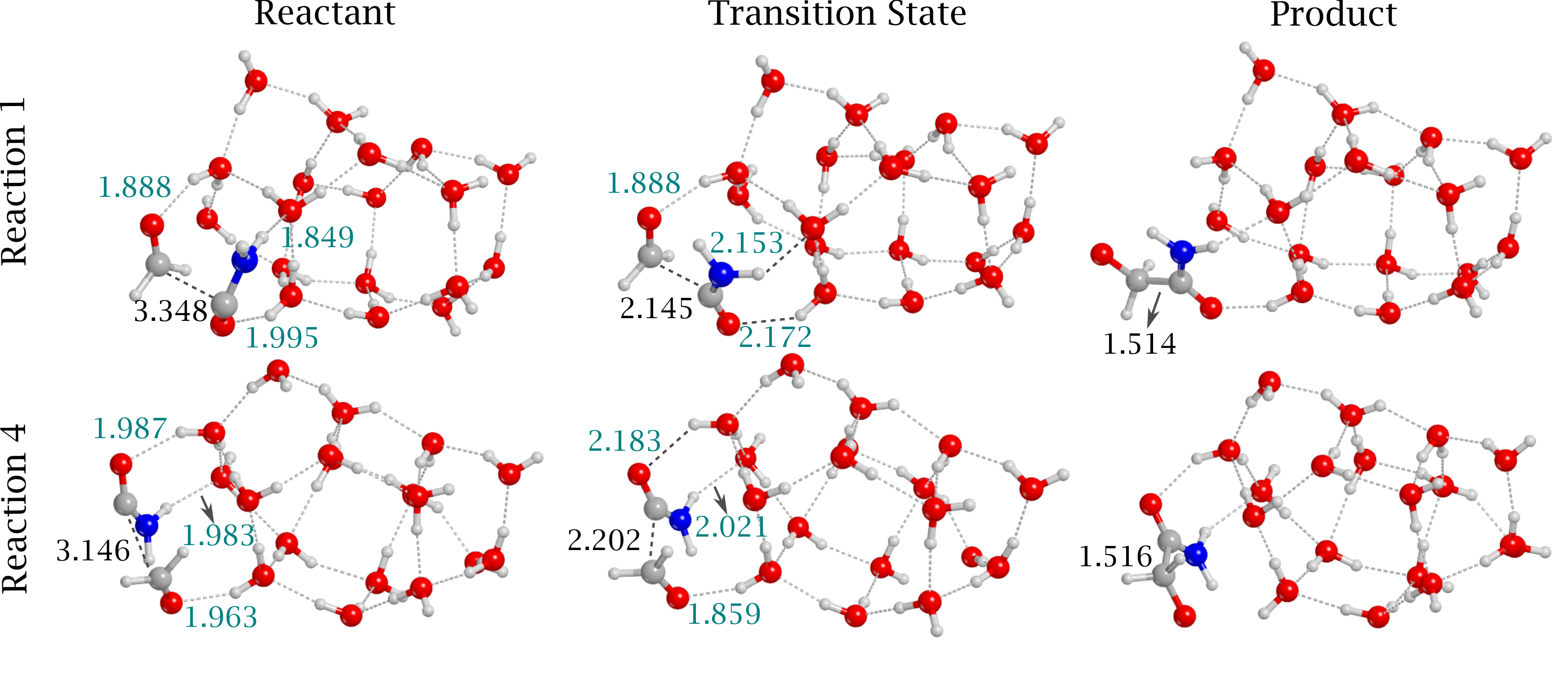}
    \caption{BHLYP-D3(BJ)/ma-def2-TZVP optimized structures of reactant, transition state, and product for Reactions 1 and 4 (those presenting the highest and the lowest energy barrier, respectively). Distances are in \r{A}, those in blue refer to H-bond lengths. Color code: red, oxygen; light gray, hydrogen; gray, carbon; blue, nitrogen. Geometries for Reaction 2 and 3 are available in the Appendix.}
    \label{fig:reaction}
\end{figure*}

For the sake of description and discussion, here we focus on the two limiting cases: Reaction 1, $\Delta$H$_{TS}$ = 26.0 kJ mol$^{-1}$, and Reaction 4, $\Delta$H$_{TS}$ = 9.6 kJ mol$^{-1}$. By giving a look at their reactant and TS structures (shown in Figure \ref{fig:reaction}), it is possible to identify an atomistic reason for this difference. In Reaction 1, the adsorption geometry of \ce{H2CO} is essentially unperturbed by the forthcoming formation of the \ce{C-C} bond (in contrast to what we observe in the product, where it is not preserved), while the orientation of \ce{NH2CO^.} with respect to the W18 cluster changes drastically, causing an elongation of the H-bonds accepted and received by the species in the TS structure. In Reactions 2 and 3, the structural changes and, therefore, the energetics are very similar to Reaction 1. Overall, these conditions are responsible for the increase of the energy barrier in ice surface reactions compared to the gas-phase one, which is characterized by $\Delta$H(0)$_{\text{TS}}$ = 16.3 kJ mol$^{-1}$.
In contrast, in Reaction 4, the H-bond interactions keeping the reactants adsorbed on the surface are slightly weaker than those of Reaction 1, explaining the lower BE. Additionally, the rearrangement of \ce{H2CO} and \ce{NH2CO^.} in the TS structure causes the strengthening of the H-bond interaction between \ce{H2CO} and W18 (which shortens by 0.1 \AA) and the interaction established by the amino group of \ce{NH2CO^.} is also preserved. The only H-bond negatively affected by the reaction is that accepted by the \ce{NH2CO^.}. Such conditions lower the energy barrier of Reaction 4 to 9.6 kJ mol$^{-1}$, thus making it more preferred than the gas-phase process. Remarkably, the geometrical features of the products obtained from Reactions 1 to 4 already suggest that the successive hydrogenation step will form \textit{anti}-glycolamide. 

\begin{table*}[htb!]
\centering
\caption{Summary of the energetic data (in kJ mol$^{-1}$) computed for the adsorption complexes chosen as reactants. 
\label{tab:reaction}}
\setlength{\tabcolsep}{11pt}%
\begin{tabular}{@{}ccccccc}
\toprule
Reaction & \ce{H2CO}\tablefootmark{a} & \ce{NH2CO^.}\tablefootmark{a} & BE$_{\text{R}}$\tablefootmark{b} & $\Delta$H(0)$_{\text{TS}}$\tablefootmark{c} & 
 $\Delta$H(0)$_{\text{R}}$\tablefootmark{d} & BE$_{\text{P}}$\tablefootmark{e}  \\
\midrule
1 & 46a & 52a & 78.0 & 26.0 &  -23.8 & 41.4 \\
2 & 32b & 19a & 73.8 & 17.9 &  -29.9 & 43.2  \\
3 & 32c & 19b & 73.9 & 23.6 &  -24.7 & 38.1  \\
4 & 52a & 46b & 67.0 &  9.6 &  -40.6 & 47.1  \\
\bottomrule
\end{tabular}
\tablefoot{
\tablefoottext{a}{The number-letter combination indicates the binding site considered (see Table~\ref{tab:be}})
\tablefoottext{b}{Total BEs of the species constituting the reactants.} \tablefoottext{c}{Energy barrier.} 
\tablefoottext{d}{Reaction energy.} 
\tablefoottext{e}{BE of the newly formed \ce{NH2COCH2O^.} product.}
}
\end{table*}

Finally, we computed the BE of the newly formed \ce{NH2COCH2O^.} intermediate, using as a reference the W18 cluster adopted for the adsorption of the reactants. The BEs range between 38.1 and 47.1 kJ mol$^{-1}$, and are thus similar to those found for \ce{NH2CO^.}. This result agrees with the fact that the interaction between \ce{NH2COCH2O^.} and the ice surface occurs mainly through the \ce{O=C-NH2} moiety.

\subsection{Kinetic analysis}

\begin{figure*}[htb]
    \centering
           \includegraphics[width=0.6\textwidth]{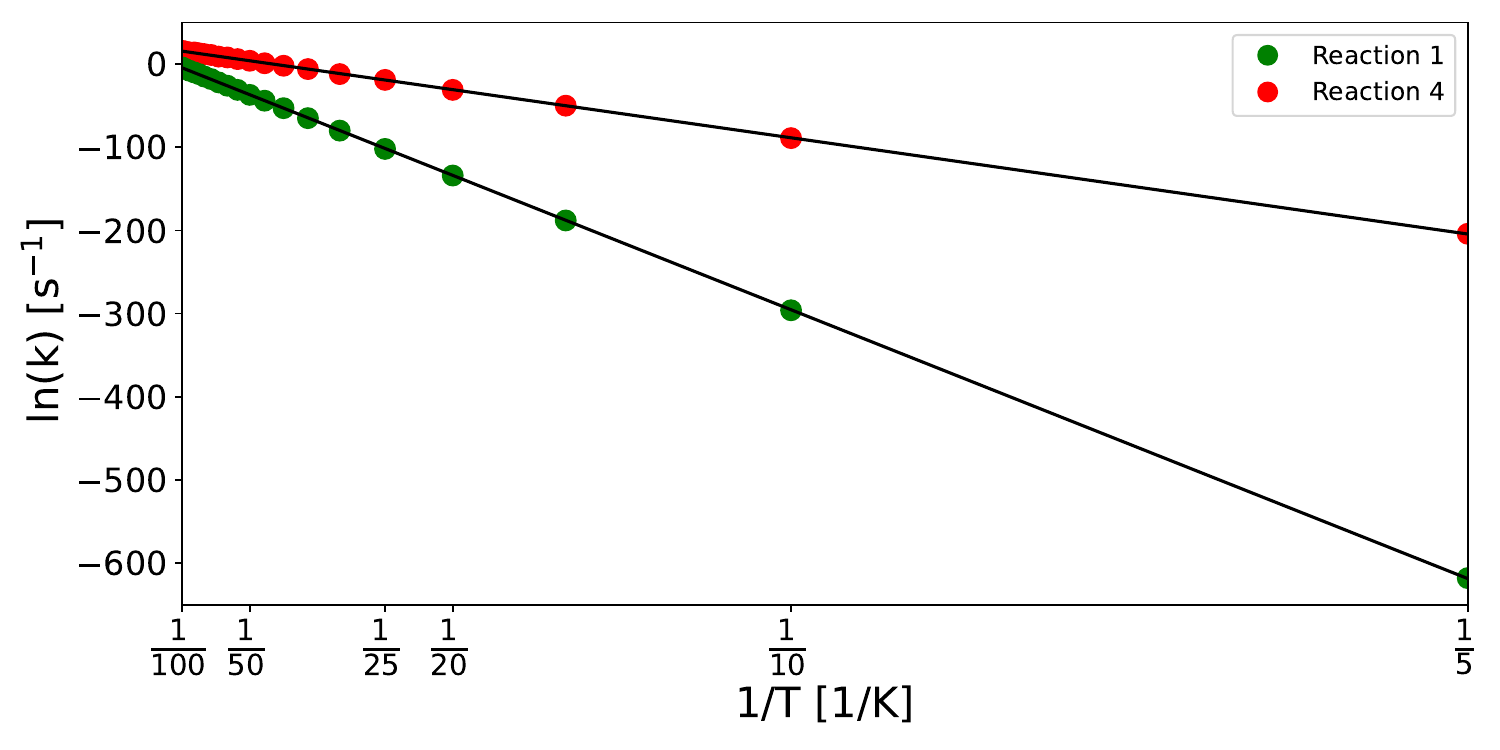} 
           \includegraphics[width=0.36\textwidth]{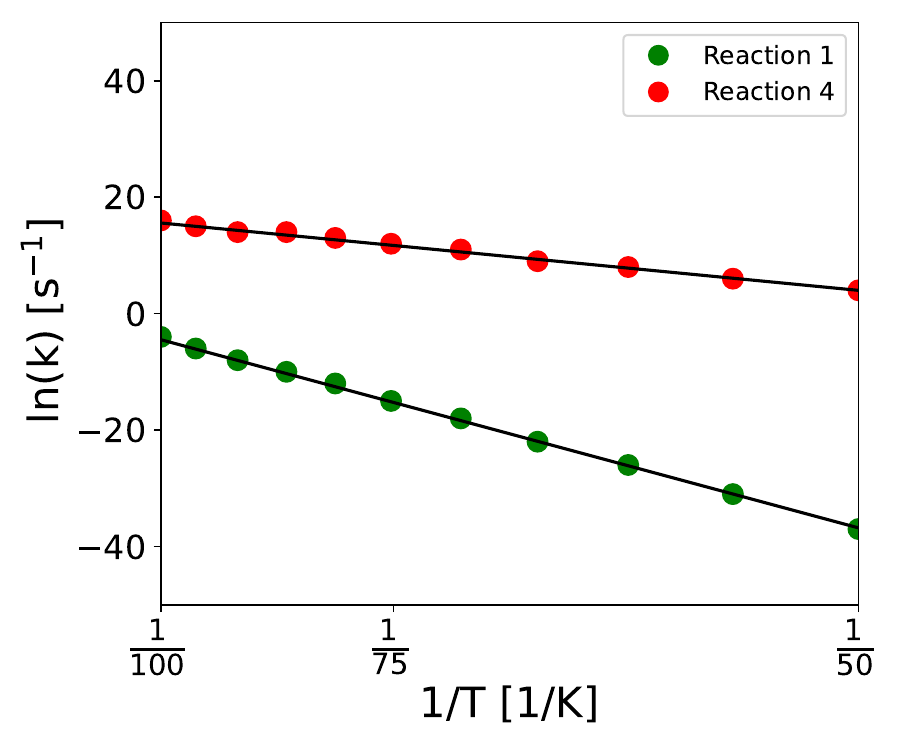} 
    \caption{Arrhenius plots for Reactions 1 and 4. The dots represent the computed kinetic constants $k$. In the left panel, the entire range of temperatures considered ($T= 5-100~\text{K}$) is shown. The right panel shows a detailed zoom-in of the Arrhenius plots in the $50-100$~K temperature range.}
    \label{fig:cinetica}
\end{figure*}

To determine whether these reactions are feasible on the surface of the icy dust grains, in the temperature range $T = 5-100~\text{K}$, we calculated the Gibbs energy barriers for Reactions 1 and 4, the unimolecular rate constants and the half-life time of the reactants (as explained in section \ref{methods}). Figure \ref{fig:cinetica} shows the Arrhenius plots for the two reactions in the temperature range considered. The resulting straight lines have a negative slope, thus pointing out that the reaction follows a classical behavior. According to our kinetic calculations, Reaction 1 is unfeasible at $T = 50 \text{K}$, at which t$_{1/2}$ = 225 Myr, and lower temperatures.

However, warmer environments are predicted to guarantee an efficient production of the glycolamide precursor. To give an example, t$_{1/2}$ is 43 s a $T = 100~\text{K}$.
On the other hand, applying Eyring's equation to Reaction 4 returns t$_{1/2}$ = 0.014~s at 50~K, which reduces to a timescale of microseconds at $T = 100~\text{K}$. Such difference between the two processes is due to the fact that the Gibbs energy enters in the exponential term of equation \ref{eq:eyring}, and consequently a small variation in the $\Delta$G$^{\ddagger}$ is responsible of a large variation in the kinetic constant and, by extension, in the half-life time.

\subsection{Hydrogenation of \texorpdfstring{\ce{NH2C(O)CH2O^.}}{NH2COCH2O}}

Hydrogen addition to the glycolamide \ce{NH2C(O)CH2O^.} radical precursor leads to the formation of the targeted final product, glycolamide. As we suspected from the previous reaction step,hydrogenation leads to the \textit{anti} conformer. The Langmuir-Hinshelwood mechanism of this final reaction involves the adsorption of the H atom in proximity of \ce{NH2C(O)CH2O^.} previously formed. To better describe this process, different orientations of the H atom with respect to \ce{NH2C(O)CH2O^.} were tested, resulting in an interaction energy $\Delta E$ of the incorporated H atom in the \ce{NH2C(O)CH2O^.}/ice system ranging from -2.1 to -6.6 kJ mol$^{-1}$ in the triplet electronic state, thus indicating a weak interaction.

\begin{table}[htb]
\centering
\caption{Hydrogenation of \ce{NH2C(O)CH2O^.} on W18 (step (\ref{eq:second})): summary of the outcomes.}
\label{tab:hydrogen}
\setlength{\tabcolsep}{7pt}%
\begin{tabular}{@{}ccc}
\toprule
Geometry\tablefootmark{a} & Barrier & Product \\
\midrule
1-H1 &barrierless& formyl formamide + \ce{H2} \\ 
1-H2 & 1.4 kJ mol$^{-1}$ & formyl formamide + \ce{H2} \\
1-H3 & 0.3 kJ mol$^{-1}$& \textit{anti-}glycolamide  \\
1-H5 &barrierless& \textit{anti-}glycolamide  \\
\midrule
4-H1 &barrierless& \textit{anti-}glycolamide  \\
4-H3 & 2.3 kJ mol$^{-1}$ & \textit{anti-}glycolamide  \\
4-H5 & 3.6 kJ mol$^{-1}$ & \textit{anti-}glycolamide  \\
4-H6 &barrierless& \textit{anti-}glycolamide  \\
4-H7 & 1.5 kJ mol$^{-1}$ & \textit{anti-}glycolamide \\
\bottomrule
\end{tabular}
\tablefoot{\tablefoottext{a}{Geometries are shown in the Appendix.}}
\end{table}

\begin{figure}[htb]
    \centering
    \includegraphics[width=0.85\linewidth]{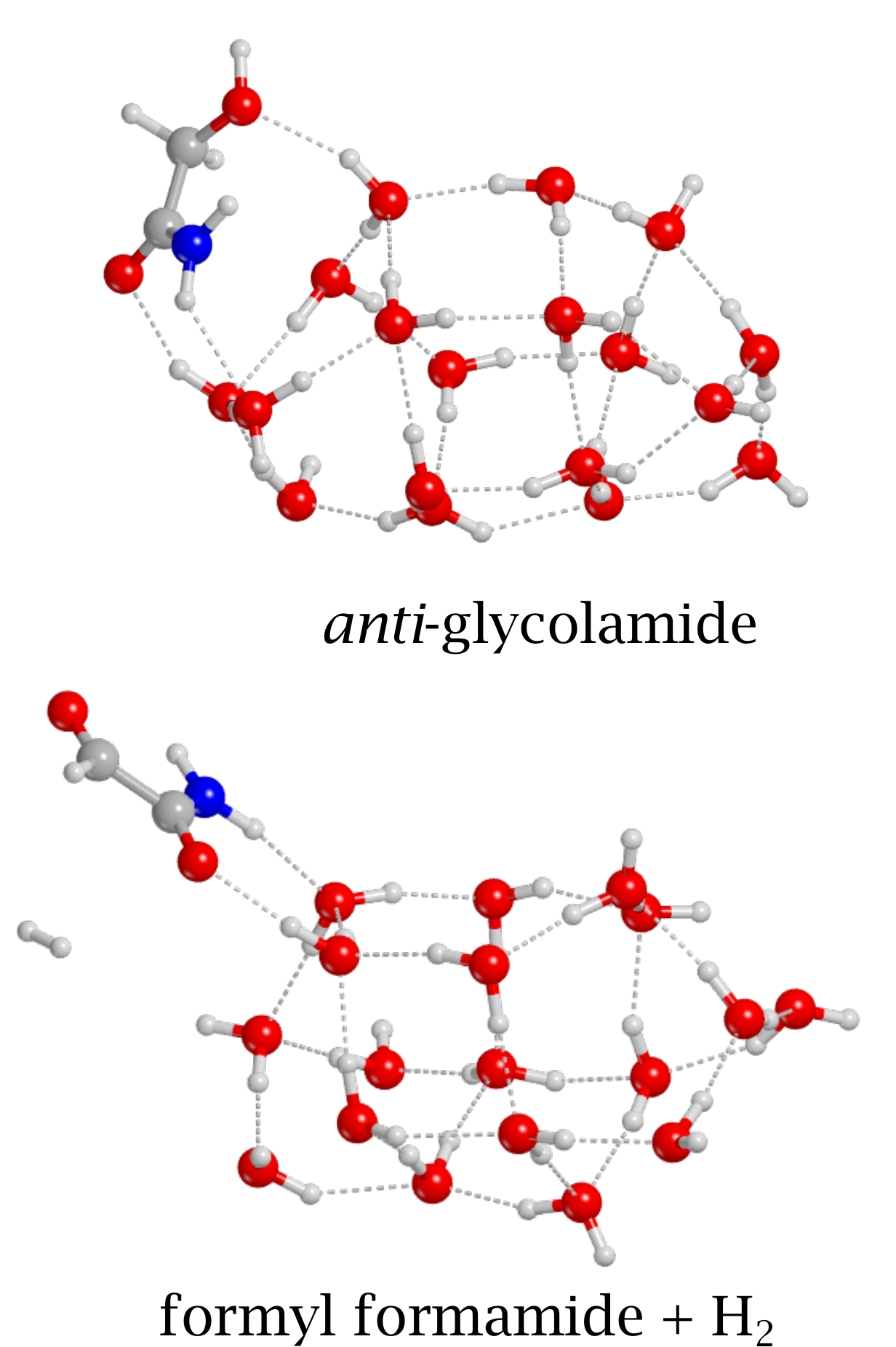}
    \caption{Hydrogenation of \ce{NH2C(O)CH2O^.} products. Top panel: the addition of atomic H to the \ce{CH2O^.} moiety leads to the formation of glycolamide. Bottom panel: the H abstraction of the \ce{CH2O^.} moiety  promotes the formation of formyl formamide (\ce{HCOCONH2}) and \ce{H2}.}
    \label{fig:hydrogen}
\end{figure}

The optimization of the \ce{NH2C(O)CH2O^./ice + H^.} complex in the biradical singlet state returned three possible scenarios (summarized in Table \ref{tab:hydrogen}): (i) the H addition forms spontaneously the \ce{O-H} chemical bond during the optimization process, thus forming glycolamide in a barrierless way; (ii) the H atom causes a hydrogen abstraction from the \ce{CH2} group, leading to \ce{H2} and formyl formamide (\ce{HCOCONH2}); or (iii) no spontaneous reaction is observed. 
In this latter case, scan calculations revealed the presence of a small diffusion barrier, between 0.3 and 3.6 kJ mol$^{-1}$, which must be overcome by the H atom in order to reach the \ce{CH2O^.} moiety and react. In this case, we also noticed the formation of both glycolamide and formyl formamide + \ce{H2} (see Figure \ref{fig:hydrogen}). A complete summary of geometries for the hydrogenation step results is reported in Figure \ref{hydrogenation_SI} of the Appendix.

\section{Discussion}\label{discussion}

In this study, the formation of glycolamide on an interstellar water ice surface has been characterized theoretically. In this process, the ice acts as a reactant concentrator and we assume it also behaves as energy dissipator, given the exothermicity of the two simulated steps \cite[in agreement with the findings observed in previous investigations, such as][]{pantaleone2020,pantaleone2021,ferrero2023nh3}.

Furthermore, the ice surface is partly responsible for the magnitude of the reaction barrier. In step (\ref{eq:first}), on the one hand, the ice surface can favor the reaction by bringing the reactants closer and slightly activating them towards reaction (as in Reaction 4). On the other hand, when strong interactions are established between the adsorbates and the W18 cluster, it can work against the reaction, by preventing the reactants from changing their conformations, meet, and react (as in Reaction 1). As a result of these considerations, we have found that the formation of the \ce{NH2C(O)CH2O^.} radical can be either more or less favored with respect to the gas-phase process from an energetic point of view, with energy barriers ranging from 9.6 to 26.0 kJ mol$^{-1}$ against the value of 16.3 kJ mol$^{-1}$ for the gas-phase process.  

Due to its large $\Delta$G$^{\ddagger}$, and according to the rate constants calculated through the Eyring's equation, Reaction 1 is kinetically favored only at temperatures close to 100 K, which however pertain to the gas phase of the G+0.693-0.027 molecular cloud, where dust grains are estimated to possess a temperature around 20--30 K \citep{rodriguez2000}. Moreover, for the final product to be formed on the ice surface, we also need to consider the hydrogenation step, which will be feasible only at low temperatures, when atomic H can be adsorbed and retained on the icy surface mantle.
At variance, Reaction 4 represents the only feasible pathway at low temperatures, with a reactants' half-life time of 5 minutes at 35 K. This possibility is more consistent with a scenario in which the glycolamide precursor forms on the cold surface of dust grains. However, in our simulations, we assume that \ce{H2CO} and \ce{NH2CO^.} are close one to each other; on the real ice surface, diffusion must be accounted for, which can be a limiting step.

\ce{H2CO} was recently detected as component of the icy mantles \citep{rocha2024}, after several studies hypothesized its presence as a consequence of the hydrogenation of CO \citep[e.g.]{fuchs2009,rimola2014,simons2020} or the reaction of atomic C with \ce{H2O} ice \citep{molpeceres2021,Ferrero2024}. Additionally, the reactivity of the ubiquitous \ce{^.CN} radical with water can explain the presence of \ce{NH2CO^.}. Indeed, \ce{^.CN} is known to form easily hemibonded complexes with water ice \citep{rimola2018,martinez2024,enriqueromero2024}, from which formamide can be obtained after subsequent hydrogenation \citep{enriqueromero2024}. Given the presence of formamide in the same molecular cloud where glycolamide was detected, it is reasonable to consider that its precursor \ce{NH2CO^.} was available at some point in the environment. 
However, as mentioned above, the probability of the encountering between \ce{H2CO} and \ce{NH2CO^.}, which is is out of the scope of the present work, can limit the efficiency of the coupling reaction.

In step (\ref{eq:second}), the specific location of the H atom on the ice cluster does not seem to hinder the hydrogenation process. Even in the case that the distance between the \ce{CH2O^.} moiety and the incoming H atom is $\sim$ 6 \r{A}, the computed diffusion barrier is only 3.5 kJ mol$^{-1}$. However, the possibility for the occurrence of another reaction yielding formyl formamide and \ce{H2}, which competes with the formation of glycolamide, must be considered. Nonetheless, this competitive reaction involves the cleavage of the \ce{C-H} bond prior to the formation of a new \ce{H-H} bond, while the production of the glycolamide only requires the formation of an \ce{O-H} bond. Furthermore, the orientation of the H atom with respect to the glycolamide precursor plays a pivotal role in determining the outcome of the reaction, with the formation of \ce{NH2C(O)CH2OH} being preferred in the various cases tested. Thus, we may suppose that the formation of glycolamide represents the main channel. 
Nevertheless, a larger number of simulations, likely involving molecular dynamics, could better characterize the outcomes of the hydrogenation step as well as the fate of its products (such as the possibility for the weakly bound \ce{H2} to desorb). 

Finally, even if our simulations led to the formation of \textit{anti-}glycolamide on the surface, they can still explain the gas-phase detection of the \textit{syn} conformer toward the G+0.693–0.027 cloud \citep{rivilla2023first}.
Given that the binding energies obtained for the glycolamide on W18 cover a range of $\text{BE} = 51.0-61.9~\text{kJ mol}^{-1}$, at temperatures such as $T$ = 10--50 K, thermal desorption of this species in the gas phase is unlikely, requiring some kind of energetic process. Concurrently, an energy source that would cause the desorption of glycolamide from the ice surface will be also sufficient to induce the conversion from the \textit{anti} conformer to the \textit{syn} form. Furthermore, the interconversion might occur thanks to NIR photons, as suggested by an isolated Ar-matrix study on glycolamide \citep{lapinski2019conformational}. 
The gas-phase \textit{anti}-to-\textit{syn} interconversion barrier computed here (see Figure \ref{fig:synanti}) is $\sim$15 kJ mol$^{-1}$, which is in agreement with the literature \citep{maris2004conformational}. Such a barrier can be easily overcome once the species has enough energy to desorb from the ice. Since the transition frequency is low ($\sim$70\textit{i}--80\textit{i} cm$^{-1}$) and the interconversion process involves the motion of several heavy atoms (i.e., C and O), tunneling is not expected to play a pivotal role in the process, despite the fact that the barrier is comparable or even lower than that of imines or carboxylic acids \citep{de2022trans,de2021origin}.

Remarkably, the requirements for the desorption and subsequent interconversion of \textit{anti-}glycolamide aligns with the conditions of the G+0.693-0.027 molecular cloud, where cloud-cloud collisions can induce shock-driven desorption. Similar arguments might also help rationalize the absence of glycolamide in Sgr B2(N) and G31.41+0.31 hot molecular core \citep{sanz-novo2020,colzi2021}.  

\section{Conclusions}\label{conclu}

In this work, the formation of glycolamide (\ce{NH2C(O)CH2OH}) has been investigated by means of quantum chemical computations using an 18 \ce{H2O} molecules cluster representing the icy mantle of interstellar dust grains. \textit{Syn-}glycolamide has been detected in the G+0.693-0.027 molecular cloud, where other amides have also been observed. This molecule is expected to provide a source for more complex amino acids via its \ce{HO^.CHC(O)NH2} intermediate \citep{joshi2025identification}. Thus, glycolamide can also play a role in prebiotic chemistry.

Glycolamide synthesis has been simulated here considering a two-step mechanism, involving the reaction between \ce{H2CO} and \ce{NH2CO^.} (step 1), followed by the hydrogenation of the prior product, \ce{NH2C(O)CH2O^.} (step 2).
First, the gas-phase (i.e., in the absence of the W18 cluster) coupling between \ce{H2CO} and \ce{NH2CO^.} was simulated. This served to determine the intrinsic energetic features of the reaction (characterized by an energy barrier of about 16~kJ mol$^{-1}$) as well as to conduct a methodological benchmark study, which led to the identification of BHLYP-D3(BJ)/ma-def2-TZVP as the most suitable level to describe this reaction.

The adsorption of \ce{H2CO} and \ce{NH2CO^.} on different surface binding sites of the W18 cluster allowed us to evaluate their binding strengths on W18 and identify the most favorable \ce{H2CO} and \ce{NH2CO^.} adsorption pairs, in which the reactants are both strongly bound to the ice surface and located in close proximity, thus favoring their reaction. 
The computed reaction barriers for the \ce{NH2C(O)CH2O^.} formation ($\Delta$ H(0)$_{TS}$ = 9.6--26.0 kJ mol$^{-1}$) indicated that the ice surface can either facilitate or hinder the process compared to the gas-phase reaction. Kinetic analysis suggests that, considering the two limit situations, the reaction is either feasible at temperatures above 35~K (lowest energy barrier case) or above 100~K (highest energy barrier case), up to the desorption of the reactants.

A final, almost barrierless, hydrogenation step leads to \textit{anti}-glycolamide formation. A competing pathway yielding formyl formamide (\ce{HCOCONH2}) and \ce{H2} exists, but according to the different cases tested, it is likely less favorable than glycolamide synthesis. 
The BE of \textit{anti}-glycolamide suggests that its thermal desorption is unlikely. However, the physical conditions of the G+0.693-0.027 molecular cloud, where cloud-cloud collisions generate shock-induced desorption, enable grain mantle sputtering, which not only allows its desorption, but also cause the interconversion between the \textit{anti}- and \textit{syn-} glycolamide conformers, thus explaining the exclusive presence of the latter form in this source and its absence toward less harsh environments.

\begin{acknowledgements}
J.P. acknowledges support from the Project CH4.0 under the MUR program ``Dipartimenti di Eccellenza 2023-2027” (CUP: D13C22003520001) and from the Italian Space Agency (Bando ASI Prot. n. DC-DSR-UVS-2022-231, Grant no. 2023-10-U.0 MIGLIORA). 
S.A. acknowledges the COST Action CA21101 ``COSY - Confined molecular systems: from a new generation of materials to the stars’’ for a STSM. H.Y. acknowledges the `` Marco Polo'' fellowship granted by the Department of Chemistry ``Giacomo Ciamician'', University of Bologna. In Bologna, this work has been supported by MUR (PRIN Grant Numbers 202082CE3T, P2022ZFNBL and 20225228K5) and by the University of Bologna (RFO funds). 
A.R. acknowledges financial support from the European Union’s Horizon 2020 research and innovation programme from the European Research Council (ERC) for the project ``Quantum Chemistry on Interstellar Grains” (QUANTUMGRAIN), grant agreement No. 865657, and from the Spanish MICINN for funding the projects PID2021-126427NB-I00 and CNS2023-144902. A.R. gratefully acknowledges support through 2023 ICREA Award. The authors thankfully acknowledge the supercomputational facilities provided by CSUC.
\end{acknowledgements}


%
\bibliographystyle{aa} 
\bibliography{glicolamide} 
\onecolumn

\begin{appendix} 

\section{Benchmark of gas-phase glycolamide formation}

In the following figures, the color code employed is: red for oxygen, light gray for hydrogen, gray for carbon, and blue for nitrogen.

\begin{figure*}[hb!]
    \centering
    \includegraphics[width=0.85\textwidth]{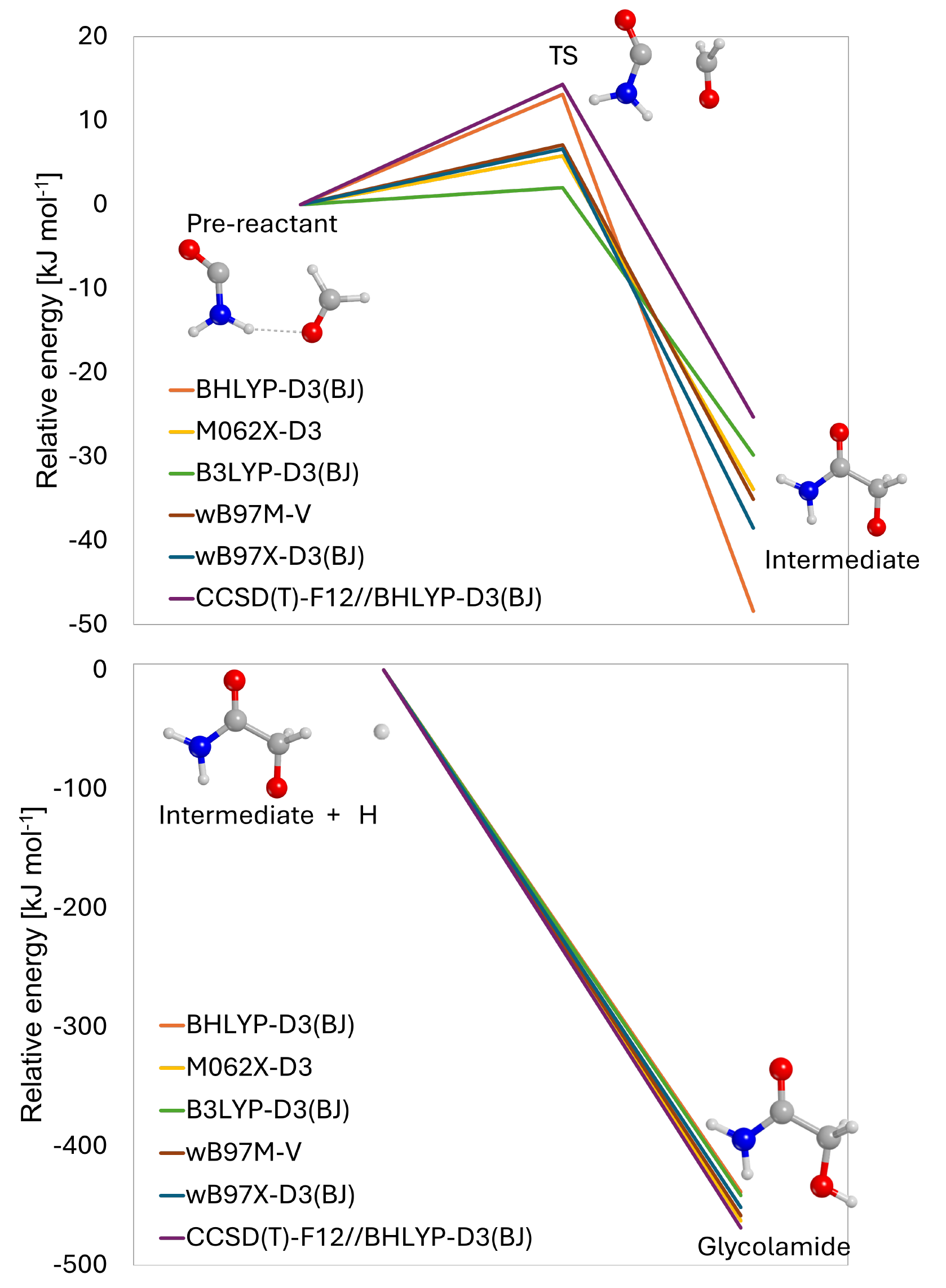}
    \caption{Relative energies of the two reaction steps required to yield glycolamide. For the sake of clarity, given that the energies computed at the CCSD(T)//DFT theory level are very similar, we adopted CCSD(T)//BHLYP-D3(BJ) as the reference in the plot. Energies are given in kJ mol$^{-1}$.}
    \label{benchmark_plot}
\end{figure*}

\section{Adsorption structures of
\texorpdfstring{\ce{H2CO}}{H2CO} on the W18 cluster.}\label{H2COad}

The XYZ files corresponding to the structures depicted in the following appendixes are freely available in Zenodo at:\\ \url{https://doi.org/10.5281/zenodo.14850268}

\begin{figure*}[hb!]
    \centering
    \includegraphics[width=\textwidth]{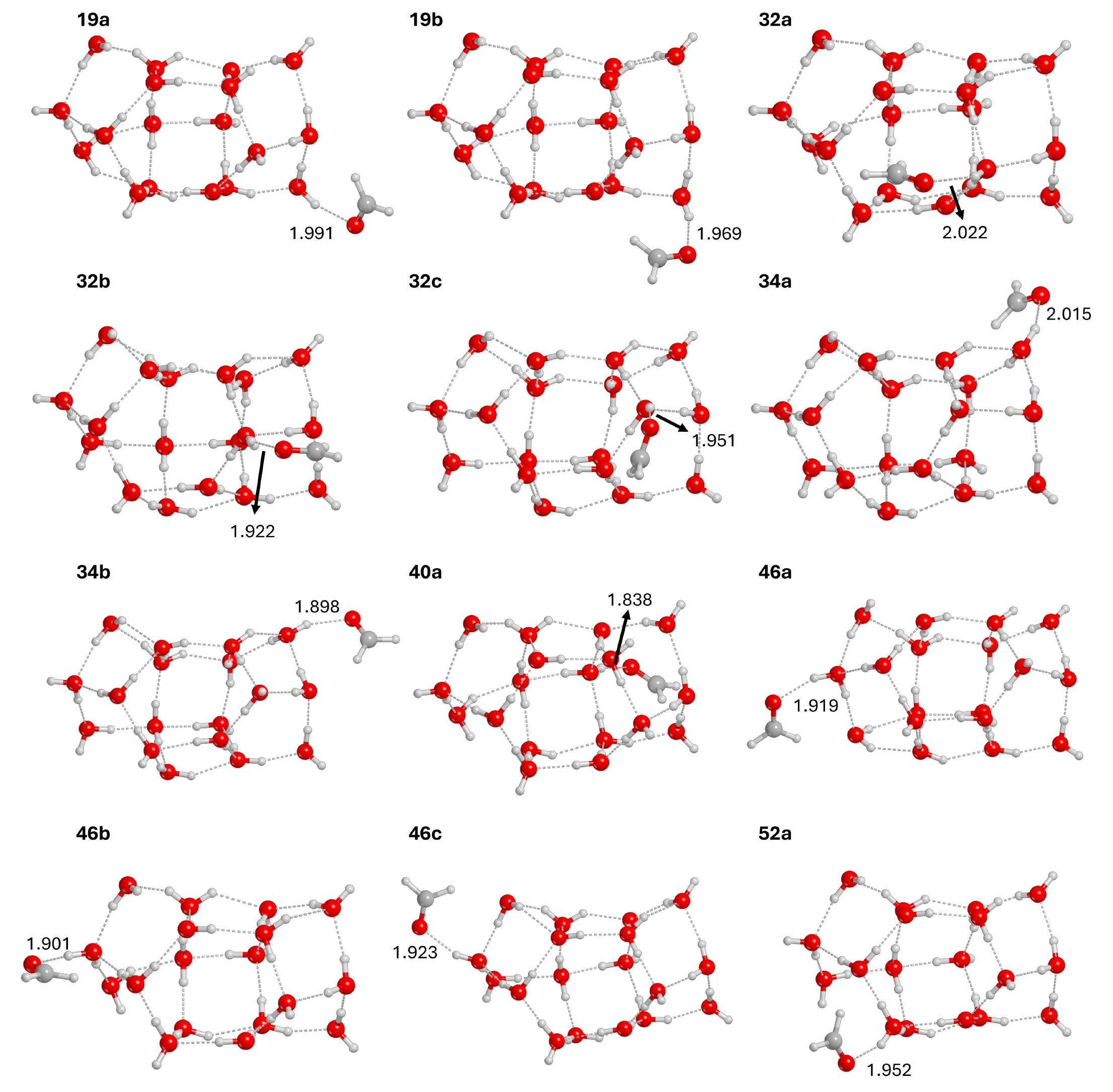}
    \caption{BHLYP-D3(BJ)/ma-def2-TZVP optimized adsorption structures of \ce{H2CO} on the W18 cluster. Distances are given in \r{A}.}
    \label{fig:adsH2CO}
\end{figure*}

\clearpage

\section{Adsorption structures of \texorpdfstring{\ce{NH2CO^.}}{NH2CO} on the W18 cluster.}\label{NH2COad}
\begin{figure*}[hb!]
    \centering
    \includegraphics[width=\textwidth]{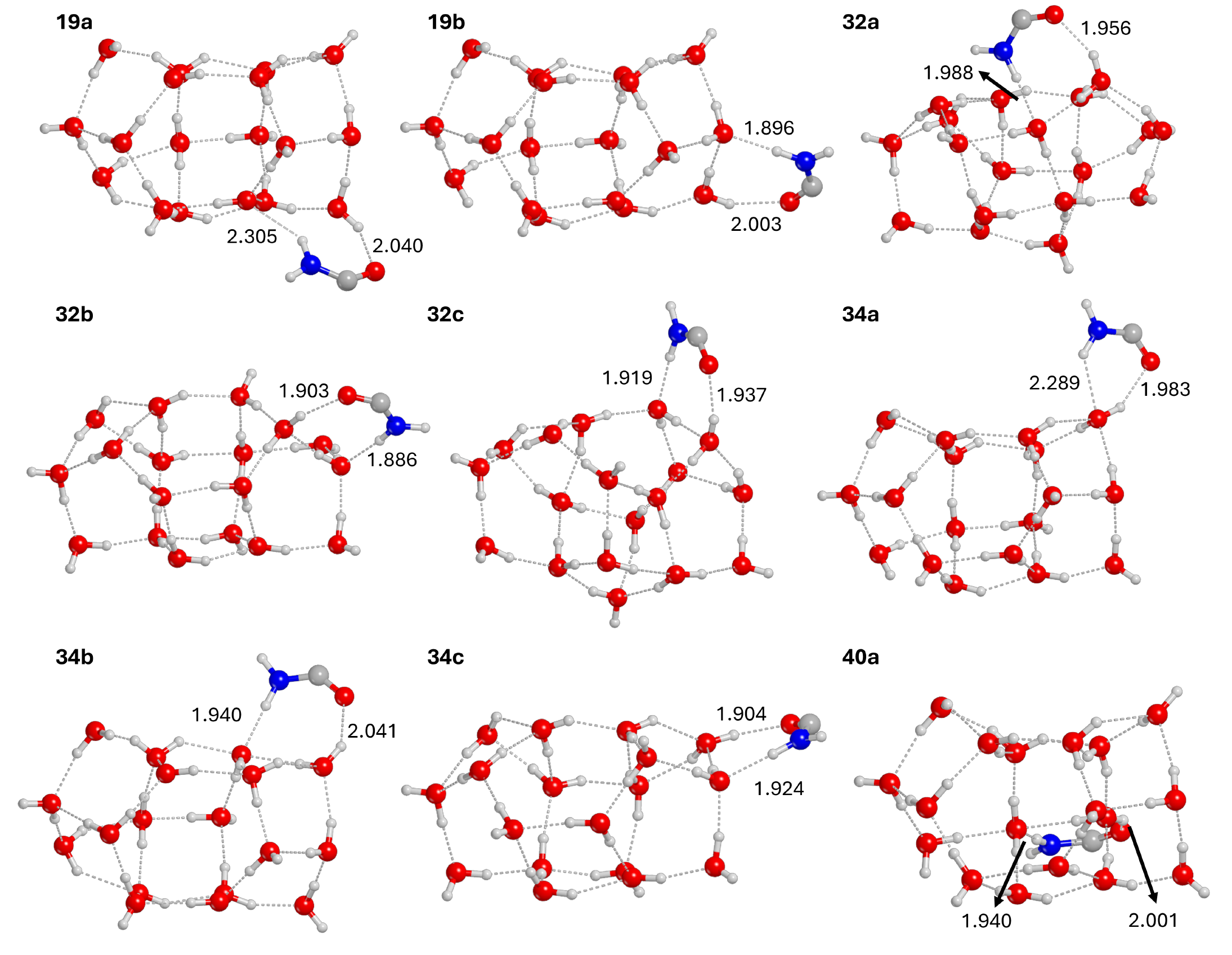}
    \caption{BHLYP-D3(BJ)/ma-def2-TZVP optimized adsorption structures of \ce{NH2CO^.} on the W18 cluster. Distances are given in \r{A}.  }
    \label{fig:adsNH2CO}
\end{figure*}

\begin{figure*}[hb!]
    \centering
    \includegraphics[width=\textwidth]{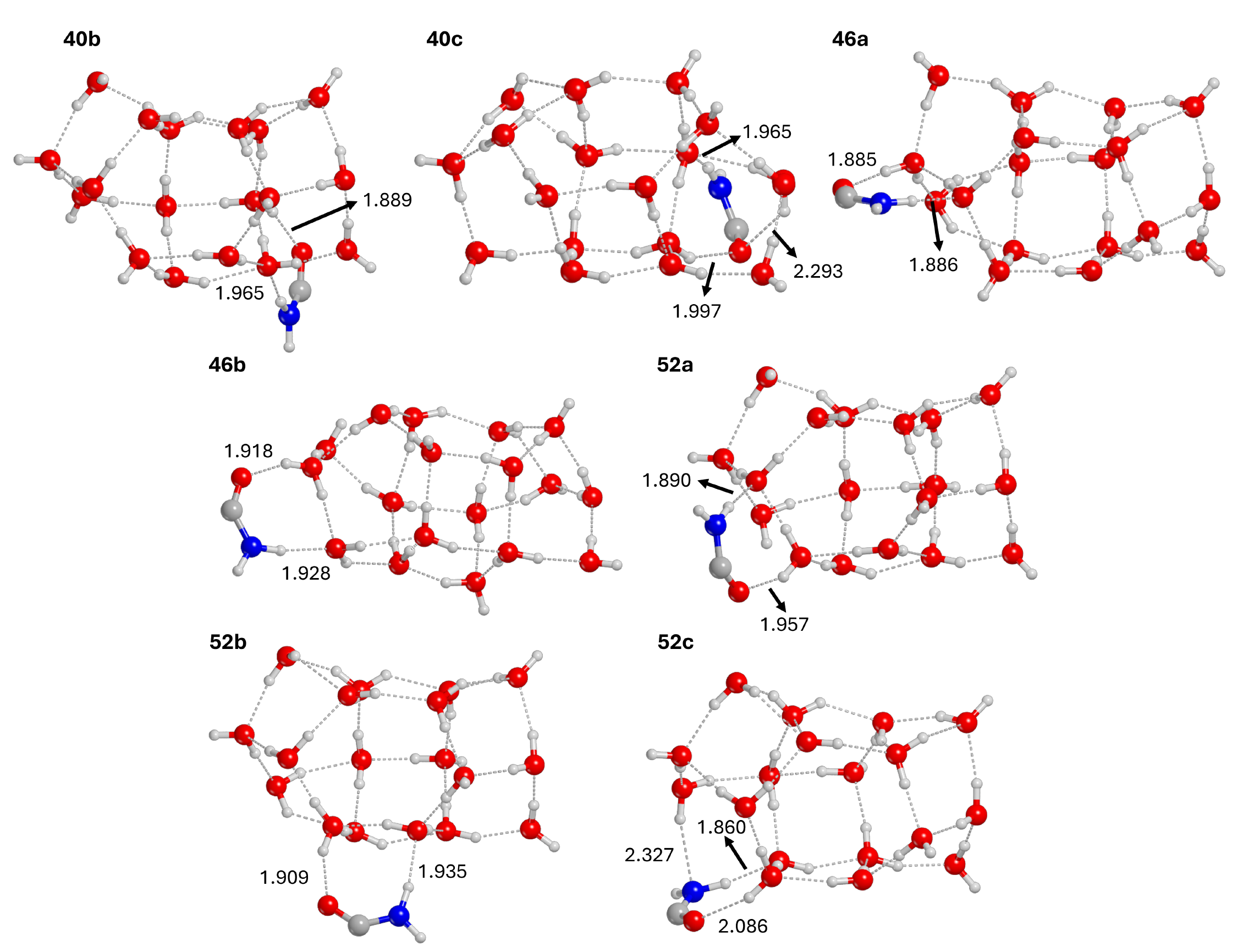}
    \caption{BHLYP-D3(BJ)/ma-def2-TZVP optimized adsorption structures of \ce{NH2CO^.} on the W18 cluster. Distances are given in \r{A}.}
    \label{fig:ads}
\end{figure*}

\clearpage

\section{Structures of reactant, transition state and product for the reaction investigated on the W18 cluster.}\label{tabrection_SI}
\begin{figure*}[hb!]
    \centering
    \includegraphics[width=\textwidth]{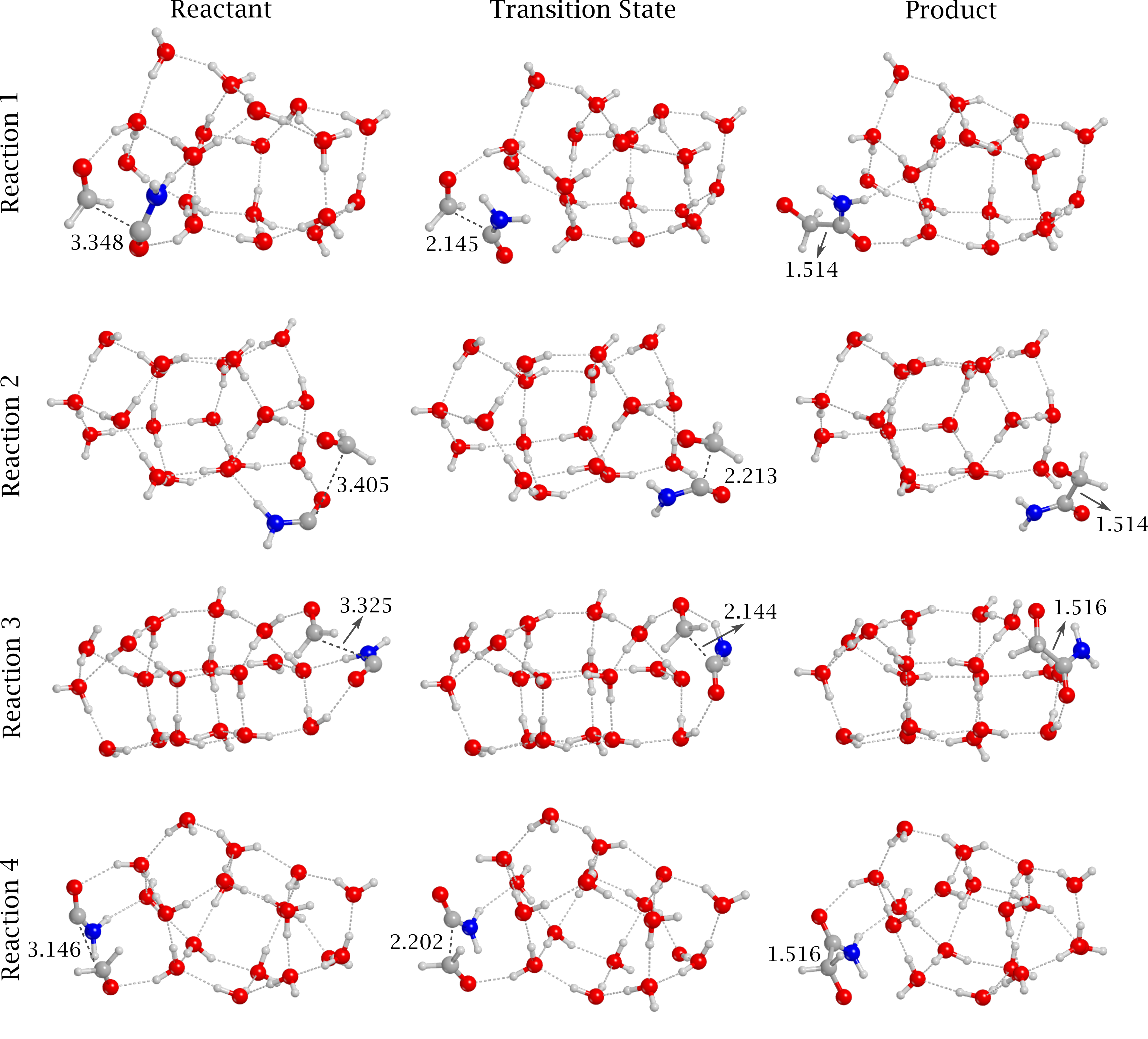}
    \caption{BHLYP-D3(BJ)/ma-def2-TZVP optimized structures of reactant, transition state, and product for Reactions 1, 2, 3, and 4. Distances are given in \r{A}.}
    \label{fig:reaSI}
\end{figure*}

\clearpage 

\section{Hydrogenation products}\label{hydrogenation_SI}
\begin{figure*}[hb!]
    \centering
    \includegraphics[width=\textwidth]{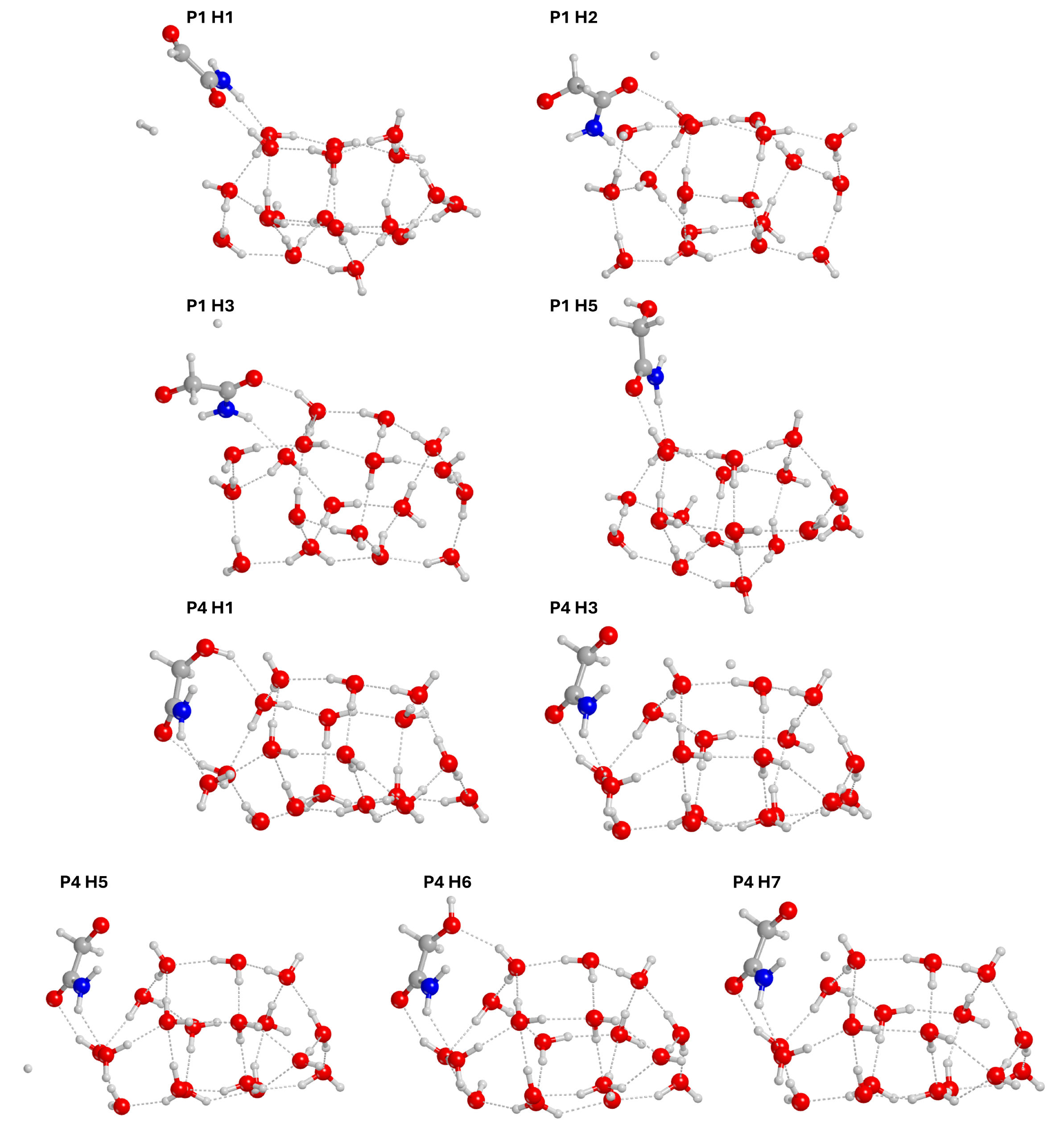}
    \caption{Different results of the hydrogenation step, forming glycolamide or formyl formamide, as well as a non reactive adduct. Distances are given in \r{A}.}
    \label{fig:hydrSI}
\end{figure*}
\end{appendix}

\end{document}